\documentclass[prd,eqsecnum,showpacs,aps,nofootinbib]{revtex4}
\usepackage{latexsym}
\usepackage{graphicx}
\begin{document}
\title{Probing violation of the Copernican principle \\via the integrated 
Sachs-Wolfe effect}
\author{Kenji Tomita }
\affiliation{Yukawa Institute for Theoretical Physics, 
Kyoto University, Kyoto 606-8502, Japan}
\author{Kaiki Taro Inoue}
\affiliation{Department of Science and Engineering, Kinki University,
Higashi-Osaka, 577-8502, Japan}  
\date{\today}

\begin{abstract}

Recent observational data of supernovae 
indicate that we may live in an underdense region, which challenges 
the Copernican principle. We show that 
the integrated Sachs-Wolfe (ISW) effect is an excellent 
discriminator between anti-Copernican inhomogeneous models and the 
standard Copernican models.  As a reference model, we consider 
an anti-Copernican inhomogeneous model that consists of two 
inner negatively curved underdense regions and an outer flat
Einstein-de Sitter region. We assume that these regions are 
connected by two thin-walls at redshifts $z = 0.067$ and $z=0.45$. 
In the inner two regions, the first-order ISW effect 
is dominant and comparable to that in the concordant flat-$\Lambda$ 
models. In the outer Einstein-de Sitter region,  
the first-order ISW effect vanishes but the second-order
ISW effect plays a dominant role, while the first-order ISW effect is
dominant in the flat-$\Lambda$ models at moderate redshifts. This
difference can discrimate the 
anti-Copernican models from the concordant flat-$\Lambda$ model. 
At high redshits, the second-order ISW effect is dominant both in
our inhomogeneous model and the concordant model.
In the outer region, moreover, the ISW effect due to large-scale
density perturbations with a present matter density contrast  
$\epsilon_{m0} \ll 0.37$ is negligible, while 
the effect due to small-scale density perturbations (such as clusters of
galaxies, superclusters and voids) with $\epsilon_{m0} \gg 0.37$ would
generate 
anisotropies which are larger than those generated by the 
ISW effect in the concordant model.
\end{abstract}
\pacs{98.80.-k, 98.70.Vc, 04.25.Nx}

\maketitle


\section{Introduction}
\label{sec:level1}

Assuming a uniform distribution of matter on large scales,
the observed data of high-redshift type Ia 
supernovae(SNIa)\cite{schm,ries1,ries2,perl} point to 
$\Lambda$-dominated flat Friedmann-Robertson-Walker (FRW) models. 
The darkness of the SNIa is reduced to accelerating expansion
of the universe due to a positive $\Lambda$ term.

These $\Lambda$-dominated FRW models are 
consistent also with the observed data of 
temperature anisotropy in the Cosmic Microwave
Background (CMB) radiation\cite{map,spg}, except for 
the low-multipole components\cite{olv, cont}.  
Moreover the observed correlation between the CMB and
large-scale structure supports 
these $\Lambda$-dominated models, which can generate 
anisotropies due to the first-order(linear) 
ISW effect\cite{bou,turok,granett}. 

On the other hand, alternative inhomogeneous models that can 
explain the SNIa data without introducing a cosmological constant
$\Lambda$ have been independently proposed by C${\rm \acute{e}}$l${\rm 
\acute{e}}$rier\cite{cele}, Goodwin et al.\cite{good} and
Tomita\cite{toma,tomb,lvm,tom} and subsequently studied by several 
authors\cite{iguchi,aln,bis,alex}. It turned out that some 
inhomogeneous cosmological models 
with an inner large-scale underdense region
(which we called {\it a local void} in our previous works) with 
a small Hubble constant ($h\approx 0.5$) in the outer flat 
region can also explain the CMB data\cite{aln,blan,hunt,alex} 
as well as the SNIa data. In these models, the 
cosmological Copernican principle is violated since 
we need to live near the center of an underdense region. 

However, recent observational
studies such as the baryon acoustic oscillations
(BAO)\cite{eisens,seo,perc1,perc2,perc3,gazt,zibin}, the kinematic
Sunyaev-Zeldovich effect either from clusters \cite{bellido2} or 
reionized regions\cite{cald} put stringent constraints on 
these anti-Copernican models.
As a result, models with a local void on 300 Mpc 
scales seem to be ruled out. At
the moment, we need to consider inhomogeneous models with a local void on
Gpc scales so that the constraints from BAO at epochs of $z \le 0.45$
may be avoided. Recently several Gpc void models have been studied by
Clifton et al.\cite{clif} and Garc\'ia-Bellido and Haugb${\rm
\o}$lle\cite{bellido1}.     
 
In this paper we study the ISW effect\footnote{In this paper, 
``the ISW effect'' means redshift/blueshift of the CMB photons
due to time-evolving first-order or second-order metric perturbations. 
} in flat FRW models 
with an inner underdense region on Gpc scales
based on previous results\cite{tom1,tom2,tom3,is1,is2,ti}. 
Then we compare it with the ISW effect in the concordant flat FRW model
with a cosmological constant $\Lambda$. 
As we shall show, the ISW effect will be 
an excellent discriminator between our anti-Copernican models and the 
standard concordant Copernican model.
In \S 2, we present our inhomogeneous 
cosmological model with inner underdense regions and in \S 3 we derive
analytic formulae for calculating the ISW effect in the inner 
and outer regions and we discuss the property of temperature
anisotropy due to the ISW effect 
in our models and the concordant model. 
\S 4 is dedicated to concluding remarks.
In what follows, we use the units of $8\pi G = c = 1$.
For spatial coordinates, we use Latin subscripts 
running from 1 to 3. 

\section{A cosmological model with inner underdense regions}
\label{sec:level2}

Our inhomogeneous anti-Copernican models without a cosmological 
constant $\Lambda$ consist of two inner
underdense regions (I and II) and an outer flat region (III). The
former regions are described by negatively 
curved FRW models ($\Omega_{I0} = 0.3$ and
$\Omega_{II0} = 0.6$) and the outer region is 
by the Einstein-de Sitter model (EdS)($\Omega_{III0} = 1$). Here in
these regions we use homogeneous models locally because the ISW effect
can be treated only in homogeneous models at present. We assume that these
regions are connected by 
two infinitesimally thin walls at redshifts $z = 0.067$ and $0.45$
corresponding 
to the boundary between I and II and the boundary between II and III,
respectively. The latter redshift value $0.45$ corresponds to 
a $\sim$ Gpc radius of the spherical underdense region. 
The Hubble constants $H_{I0},
H_{II0}$ and $H_{III0}$ in these regions satisfy
a relation $H_{I0} \ge H_{II0} \ge H_{III0}$. Here we consider the
following two cases:

%
\begin{eqnarray}
  \label{eq:m0}
{\rm case}\ 1.&&\ H_{I0} =60,\quad H_{II0} =50, \quad H_{III0} = 50 
\ {\rm km/s/Mpc},\cr
{\rm case}\ 2.&&\ H_{I0} =70,\quad H_{II0} =55, \quad H_{III0} = 50 
\ {\rm km/s/Mpc},
\end{eqnarray}


\noindent
where $H_{III0} (= 50)$ stands for the value necessary for the observed
CMB anisotropies in the EdS model, $H_{I0} (= 70)$ in case 2 is the standard
value in the local measurement, and the case 1 with smaller $H_{I0}$ and
$H_{II0}$ is taken so as to consider a stringent observational
condition which is given by the kinematic Sunyaev-Zeldovich
effect\cite{bellido2}. 

If we regard the outer region as the background, the inner region can be
interpreted as a local inhomogeneity 
and has an optical influence on the temperature of
CMB radiation. If the observer is exactly at the center, the influence
is isotropic, but if he is off-center, it brings
dipole, quadrupole and the other multipole anisotropies. These
anisotropies have already been analyzed and discussed in previous
papers\cite{tomdip,moff,alndp}. In what follows, we study the ISW effect
due to small-scale density perturbations (of a simple spherical top-hat
type) in the three regions in the inhomogeneous model.  

In spherical coordinates $(r,\theta,\phi$), 
the background metric of a constantly negatively curved spacetime in the 
inner regions I and II containing pressureless
matter with a matter density $\rho$ is given by    
\begin{eqnarray}
  \label{eq:m1}
 ds^2 &=& a^2(\eta) (-d\eta^2 + dl^2), \cr
 dl^2 &\equiv& \gamma_{ij} dx^i dx^j = dr^2 + \sinh^2 (r) (d\theta^2 +
\sin^2 \theta ~d\phi^2), 
\end{eqnarray}
and 
\begin{eqnarray}
  \label{eq:m1a}
 a(\eta) &=& a_* (\cosh \eta -1), \quad t = a_* (\sinh \eta - \eta), \cr
 \rho a^2 &=& 3 [(a'/a)^2 -1] = 6/(\cosh \eta -1),  
\end{eqnarray}
where $t$ and $\eta$ are the cosmic time and the conformal time. 
Prime $'$ represents $d/d \eta$ and $a_*$ is a constant.
The Hubble parameter, the density parameter and the 
redshift are 
\begin{equation}
  \label{eq:m2}
H_\alpha \equiv a'/a^2 = {\sinh \eta_{\alpha} \over a_{\alpha *}
(\cosh \eta_{\alpha} -1)^2}, \quad 
\Omega_{\alpha m} = {2 \over
\cosh \eta_{\alpha} +1}, \quad z_{\alpha}+1 
= {\cosh \eta_{\alpha 0} -1 \over \cosh \eta_{\alpha} -1},
\end{equation}
where $\alpha$ is I or II, $\eta_{\alpha 0}$ is the present value of the
conformal time  $\eta$ and for $a_{\alpha 0} \equiv a(\eta_{\alpha 0})$ we have
$a_{\alpha 0} H_{\alpha 0} =\sinh \eta_{\alpha 0}/(\cosh \eta_{\alpha 0} -1)$ and
$\Omega_{\alpha m 0} = 2/(\cosh \eta_{\alpha 0 } +1)$. The constant $a_*$
for region I or II is given by
\begin{equation}
a_{\alpha *}=\frac{\Omega_{\alpha m 0}}{ 2 (1-\Omega_{\alpha m 0})^{3/2}
 H_{\alpha 0}},
\end{equation}
where $H_{\alpha 0}=H_{\alpha}(\eta_{\alpha 0})$.

In the outer region III containing pressureless matter, the space-time 
metric is 
\begin{equation}
  \label{eq:m3}
ds^2 = a^2(\eta) [-d\eta^2 + \delta_{ij} dx^i dx^j],
\end{equation}
where $a(\eta) \propto \eta^2$. The Hubble parameter, the density 
parameter and the redshift are $H_{III} \equiv a'/a^2 = 2/(\eta a), 
\ \Omega_{IIIm} = 1$, and $z+1 = (\eta_{III0}/\eta)^2$. Here for $a_0 = 
a(\eta_{III0})$ we have as $a_0 H_{III0} = 2/\eta_{III0}$.

For comparison, we consider a concordant flat FRW model with a cosmological
constant $\Lambda$. The metric (\ref{eq:m1}) is  
$dl^2 = \delta_{ij} dx^idx^j$ and the
scale factor satisfies $3(a'/a)^2 = 
(\rho_B + \rho_\Lambda), \ 6(a'/a)' = -(\rho_B -2\rho_\Lambda)a^2$,
where $\rho_B$ and $\rho_\Lambda$ are the energy 
density of matter and that of a cosmological constant $\Lambda$, 
respectively. As the model parameter of the concordant 
flat-$\Lambda$ model, we adopt
$\Omega_{m0} = 0.3$ and $H_0 = 70$ km/s/Mpc.

\section{Integrated Sachs-Wolfe effect due to density perturbations}
\label{sec:level3}

Now we consider growing mode of 
density perturbations and the integrated Sachs-Wolfe
effect in the inner and outer regions, separately.

\subsection{The inner regions I and II}
The first-order gauge-invariant growing density
perturbations $\epsilon_{mI}$ and the gauge-invariant potential
perturbation (in the growing mode) $\Phi_A (= \Phi_H)$\cite{bard} are
expressed as  
\begin{eqnarray}
  \label{eq:r1}
\epsilon_{I m} &=& -G(\eta) \Delta F \cr
\Phi_A &=& - {1\over 2}\rho a^2 G(\eta) F
\end{eqnarray}
with 
\begin{eqnarray}
  \label{eq:r2}
G(\eta) &\equiv& {6 \over \cosh \eta -1} \ \Bigl(1 - {\eta (\cosh \eta
+1) \over 2 \sinh \eta} \Bigr) + 1, \cr
&=& {1\over 10}\eta^2 (1 - {5\over 84}\eta^2 + \cdot \cdot \cdot)
\quad {\rm for} \quad \eta \ll 1,
\end{eqnarray}
where $\epsilon_{mI}$ corresponds to the density perturbations in the
comoving synchronous gauge, $\Phi_A$ is equal to the potential
perturbations $\phi^{(1)}$ in the longitudinal or Poisson gauge, and
$F$ is the potential function given as an arbitrary function of
spatial coordinates. The expression of 
$G(\eta)$ was derived from the solution shown by Lifshits and
Khalatinikov\cite{lk}. $\Delta F$ is the Laplacian
of $F$ in the space $dl^2$, that is, $\Delta F = F_{|i}^{|i}$, where
$|i$ is covariant derivatives in the three dimensional space with
$\gamma_{ij}$. So we obtain from Eqs.(\ref{eq:m1a}) and (\ref{eq:r1}) 
\begin{equation}
  \label{eq:r3}
\phi^{(1)} = - {3 G(\eta) \over \cosh \eta -1} F.
\end{equation}
The first-order gauge-invariant temperature fluctuation due to the
linear ISW effect is expressed as 
\begin{equation}
  \label{eq:r4}
\Delta T^{(1)}/T = -2 \int^{\lambda_e}_{\lambda_o} d\lambda 
\ {\phi'}^{(1)}
\end{equation}
in the Poisson gauge, where the prime is $\partial/\partial \eta$ and
$\lambda$ is the affine parameter along the light path. $\lambda_e$
and $\lambda_o$ are the emitter's and observer's values at the
decoupling and present epochs, respectively.

\begin{figure}[t]
\caption{\label{fig:rs2} The matter density 
contrast for a top-hat type spherical void.}
\includegraphics[width=8cm]{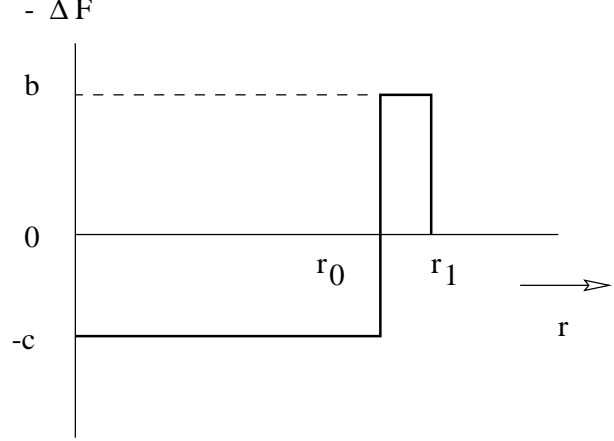}
\end{figure}

In this paper we consider a simple spherical top-hat type of
compensated density perturbation following our previous paper\cite{ti},
whose spatial size is much
smaller than the horizon size. The spatial variation of the density 
perturbation is schematically shown in Fig.\ref{fig:rs2}. We consider
the CMB photon paths
passing through the center of the spherical perturbation. When the
epoch in the center of the perturbation is $\eta$, the integral of
$\phi^{(1)}$ along the light path reduces approximately to
\begin{equation}
  \label{eq:r5}
(\Delta T/T)_\alpha = \Delta T^{(1)}/T = -{6 (\epsilon_{\alpha m 0})_c \over
G(\eta_0) (\cosh \eta -1)^3} [-(14+\cosh \eta)\sinh
\eta + 3\eta (2\cosh \eta +3)] \int^{\lambda_e}_{\lambda_o} d\lambda
\ F/c,
\end{equation}
where $\alpha$ is I or II, $(\epsilon_{\alpha m 0})_c$ and a constant
$c$ are the central values of 
$\epsilon_{\alpha m 0}$ and $\Delta F$, respectively, and the subscript $0$
denotes the present epoch.
From integration of $F$ for the above top-hat type perturbations 
derived in the previous paper\cite{ti},  we obtain
\begin{eqnarray}
  \label{eq:r6}
(\Delta T/T)_\alpha  &=& (\epsilon_{I m 0})_c \Bigl({a_0r_1 \over
(H_{I0})^{-1}} \Bigr)^3 \theta_\alpha, \cr
\theta_\alpha &\equiv&  - {4 \over 3}{(1-\Omega_{\alpha m 0})^{3/2}
\over G(\eta_0)(\cosh \eta -1)^3} [-(14+\cosh \eta)\sinh
\eta + 3\eta (2\cosh \eta +3)] w_1 (y) \cr
&\times&  \Bigl(\epsilon_{\alpha m 0}/\epsilon_{I m 0}\Bigr)_c
\Bigl(H_{\alpha 0}/H_{I0}\Bigr)^3, 
\end{eqnarray}
where $w_1(y)$ is defined as $w_1(y) = -y \ln (1+ 1/y), \ y=b/c$ and
$r_1/r_0 = (1 + 1/y)^{1/3}$. For the value of $y$, we adopt $y = 0.5$
as an example.

\subsection{The outer region III}
The first-order ISW effect does not appear and
the second-order ISW effcet is the lowest one. The second-order
temperature fluctuations were derived in our previous paper\cite{ti} and
expressed as 
\begin{equation}
  \label{eq:r7}
\Delta T^{(2)}/T = {4\over 27}\ c^2\ (r_1)^3\ w_2(y)
(\zeta_1 + 9\ \zeta_2)', 
\end{equation}
where $w_2(y) \equiv y[1- y \ln(1+1/y)], \ (\zeta_1 + 9\ \zeta_2)' = -
(39/700) \eta$ \ for the EdS model, $r_1$ is the radius of
inhomogeneities (cf. Fig. \ref{fig:rs2}), and the central value of the
density perturbation $(\epsilon_{II m})_c$ is related to a constant $c$
as   
\begin{equation}
  \label{eq:r9}
(\epsilon_{III m})_c = - {1\over 20} \eta^2 c.
\end{equation}
Using Eq.(\ref{eq:r9}), the temperature fluctuations are expressed 
as
\begin{eqnarray}
  \label{eq:r10}
(\Delta T/T)_{III} &=& \Delta T^{(2)}/T = -{26\over 63}\
\Bigl({a_0r_1\over (H_{III0})^{-1}} 
\Bigr)^3 {{(\epsilon_{III m 0})_c}^2\over (1+z)^{1/2}} \ w_2(y), \cr
&=& (\epsilon_{I m 0})_c \Bigl({a_0r_1 \over
(H_{I0})^{-1}} \Bigr)^3 \theta_{III}, \cr
\theta_{III} &\equiv& -{26\over 63}\
 {{(\epsilon_{III m 0})_c}^2\over (\epsilon_{I m 0})_c (1+z)^{1/2}} \
 \Bigl({H_{III0} \over H_{I0}} \Bigr)^3 \ w_2(y).
\end{eqnarray}
The temperature fluctuations are negative definite. They are not
exactly observed fluctuations, because their observed values should be
the difference from the average value $\langle \Delta T^{(2)}/T
\rangle$ of the sum of the second-order 
temperature fluctuations which is caused by all possible primordial
density perturbations and renormalized into the background temperature. 
This average value is derived, taking account of power spectrum of
density perturbations, in the procedure shown in a separate
paper\cite{tompp}. So the above $(\Delta T/T)_{III}$ should be here
used to show the order of magnitude of second-order ISW effect, but
for the perturbations with large amplitudes, the above second-order
fluctuations are regarded approximately as observed values, as the
mean value can be neglected. 

\subsection{The junction condition}
The deformation of the walls brings the complicated
perturbations inside the walls and their neighborhoods, as was studied by
one of the present authors through the analysis of the junction
condition\cite{tomjunct}.  They include not only density perturbations,
but also gravitational-wave and rotational
perturbations. Gravitational-wave perturbations propagate, but their
amplitudes are very small and the contribution to density
perturbations is negligible, because of the small coupling between
them. Moreover the density and rotational perturbations caused by the 
perturbed walls
do not propagate in the present dust matter models and are constrained
in the just neighborhoods of the walls. In the most part of the I, II
and III regions, therefore, we see density perturbations which are
independent of the wall motions and were caused primordially due to the
common origin. 

Their amplitudes in the three regions were nearly equal at the 
early stages, but the present amplitudes became different, because
they had different growth rates.    
Here we neglect the above complicated perturbations inside the walls
and in their narrow neighborhoods. Then the three density perturbations
$(\epsilon_{Im0})_c, (\epsilon_{IIm0})_c$ and 
$(\epsilon_{IIIm0})_c$ included in the two equations (\ref{eq:r6})
and (\ref{eq:r10}) are related as follows. First we assume that 
$\epsilon_{Im}, \epsilon_{IIm}$ and $\epsilon_{IIIm}$ should be equal
at early epochs of equal densities with the redshifts $z_1, z_2$ and
$z_3 \gg 1$, i.e. $\epsilon_{Im}(z_1) = \epsilon_{IIm}(z_2) =
\epsilon_{IIIm}(z_3)$, where $\rho_I (z_1) = \rho_{II} (z_2) =
\rho_{III} (z_3)$. The present densities $\rho_{I0}, \rho_{II0}$ and
$\rho_{III0}$ are related as $\rho_{I0}/\rho_{III0} =
(\Omega_{I0}/\Omega_{III0})(H_{I0}/H_{III0})^2$ and
$\rho_{II0}/\rho_{III0} = 
(\Omega_{II0}/\Omega_{III0})(H_{II0}/H_{III0})^2$. Then $z_1, z_2$
and $z_3$ are related as 
\begin{eqnarray}
  \label{eq:r10a}
(1+z_3)/(1+z_1) &=& [\rho_{I0}/\rho_{III0}]^{1/3} = 
[(\Omega_{I0}/\Omega_{III0}) (H_{I0}/H_{III0})^2]^{1/3}, \cr
(1+z_3)/(1+z_2) &=& [\rho_{II0}/\rho_{III0}]^{1/3} =  
[(\Omega_{II0}/\Omega_{III0}) (H_{II0}/H_{III0})^2]^{1/3}.  
\end{eqnarray}
Taking account of the growth rates, we obtain 
$\epsilon_{Im0} = \epsilon_{Im}(z_1)
G(\eta_{I0})/G(\eta_{I1})$, $\epsilon_{IIm0} = \epsilon_{IIm}(z_2)
G(\eta_{II0})/G(\eta_{II2})$ and  $\epsilon_{IIIm0} = \epsilon_{IIIm}(z_3)
(1+z_3)$. Accordingly, we obtain 
\begin{eqnarray}
  \label{eq:r10b}
\epsilon_{IIIm0} &=& (1+z_3) \epsilon_{Im0}
G(\eta_{I1})/G(\eta_{I0}), \cr
\epsilon_{IIIm0} &=& (1+z_3) \epsilon_{IIm0}
G(\eta_{II2})/G(\eta_{II0}),
\end{eqnarray}
where $z$ and $G(\eta)$ are calculated
using Eqs.(\ref{eq:m2}) and (\ref{eq:r2}).  Here we set $z_1 = 1000$. Then
 we have $(\epsilon_{IIIm0}/\epsilon_{Im0},
\epsilon_{IIIm0}/\epsilon_{IIm0})$\ are \ $(1.65, 1.15)$ and $(1.83,
1.22)$ \ in cases 1 and 2, respectively.

\subsection{Flat-$\Lambda$ models}
For comparison, we show the first-order and second-order temperature
fluctuations in 
the flat-$\Lambda$ models with $\Omega_m + \Omega_{\Lambda} =
1$, which were derived as $(\Delta T^{(1)}/T)_{loc}$ and $(\Delta
T^{(2)}/T)_{loc}$ in our previous paper\cite{ti}. They are expressed as
\begin{eqnarray}
  \label{eq:r11}
(\Delta T^{(1)}/T)_{loc} &=& (\epsilon_{Im0})_c \Bigl({a_0r_1\over
(H_{I0})^{-1}} \Bigr)^3 \theta^{(1)}_\Lambda, \cr
\theta^{(1)}_\Lambda &\equiv& {4\over 9} \Bigl[{2\Bigl({a'\over
a}\Bigr)^2 -{a'' \over a} \over {a'\over a}P' -1} \Bigr]_0
\Bigl({a'\over a}\Bigr)_0^{-3}  \Bigl [{a'\over a}+\Bigl({a'' \over a}
- 3\Bigl({a'\over a}\Bigr)^2\Bigr) P' \Bigr]\
\Bigl({\epsilon_{m0}\over \epsilon_{Im0}}\Bigr)_c
\Bigl({H_0\over H_{I0}}\Bigr)^3 w_1(y), 
\end{eqnarray}
and 
\begin{eqnarray}
  \label{eq:r12}
(\Delta T^{(2)}/T)_{loc} &=& (\epsilon_{Im0})_c \Bigl({a_0r_1\over
(H_{I0})^{-1}} \Bigr)^3 \theta^{(2)}_\Lambda, \cr
\theta^{(2)}_\Lambda &\equiv& {16\over 27}  \Bigl[{2\Bigl({a'\over
a}\Bigr)^2 -{a'' \over a} \over {a'\over a}P' 
-1}  \Bigr]_0^2  \Bigl({a'\over a}\Bigr)_0^{-3} (\zeta_1 + 9\zeta_2)'
\Bigl({(\epsilon_{m0})_c^2 \over (\epsilon_{m0I})_c}\Bigr)
\Bigl({H_0\over H_{I0}}\Bigr)^3 w_2(y), 
\end{eqnarray}
where $\epsilon_m$ and $(\epsilon_{m0})_c$ are first-order density
perturbation and its central value at present epoch, and 
$P(\eta), Q(\eta), \zeta_1(\eta)$ and $\zeta_2(\eta)$ are
auxiliary quantities used in the previous paper\cite{ti} and their
definitions are shown in Appendix. 

\subsection{Analyses and results}
In the following, we consider the behaviors of
fluctuations due to the ISW effect 
in our inhomogeneous model in comparison with those 
in the concordant flat-$\Lambda$ model. 

To do so, we
show the amplitudes of temperature anisotropy due to the ISW
effect from a spherical compensating void/cluster with a given
comoving radius and the density contrast, represented by
the five quantities $\theta_I, \theta_{II}, \theta_{III},
\theta^{(1)}_\Lambda$ and 
$\theta^{(2)}_\Lambda$. Note that the first-order quantities $\theta_I$, $\theta_{II}$ and
$\theta^{(1)}_\Lambda$ do not depend on $(\epsilon_{Im0})_c$ and
$(\epsilon_{m0})_c$, while the second-order ones $\theta_{III}$ and 
$\theta^{(2)}_\Lambda$ are proportional to $(\epsilon_{Im0})_c$ and
$(\epsilon_{m0})_c$, respectively. Here $c$ denotes the values at the
centers of the voids/clusters. 
We assume that $(\epsilon_{m0})_c = (\epsilon_{Im0})_c$, so that the
cosmological situation in the neighborhood of our observer in the
flat-$\Lambda$ model may be equal to that in the inner region I
of the model.     

\begin{figure}[htbp]
 \begin{tabular}{cc}
  \begin{minipage}{0.5\hsize}
   \begin{center}
    \includegraphics[width=11cm,clip]{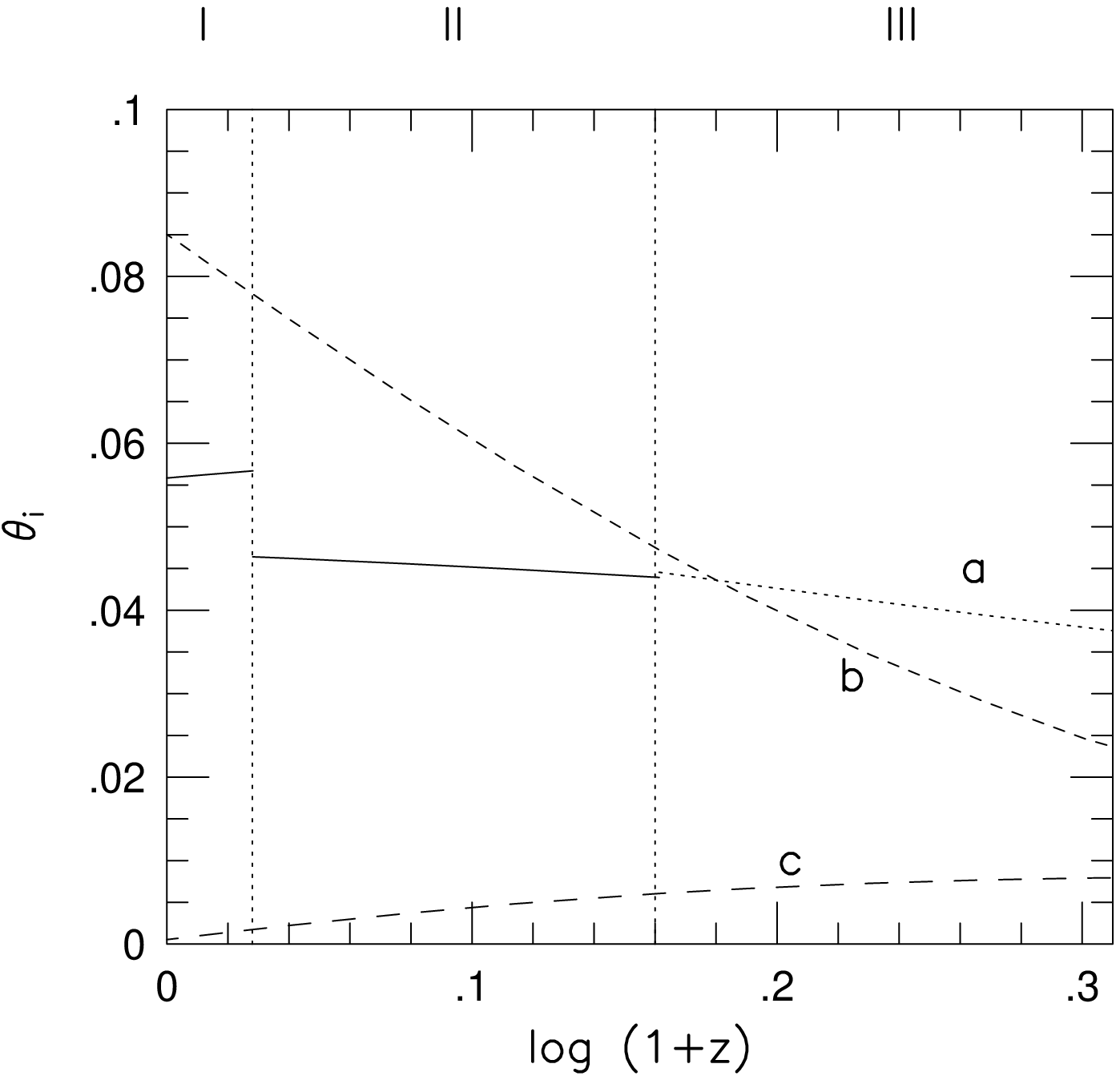}
   \end{center}
  \end{minipage}
  \begin{minipage}{0.5\hsize}
   \begin{center}
    \includegraphics[width=11cm,clip]{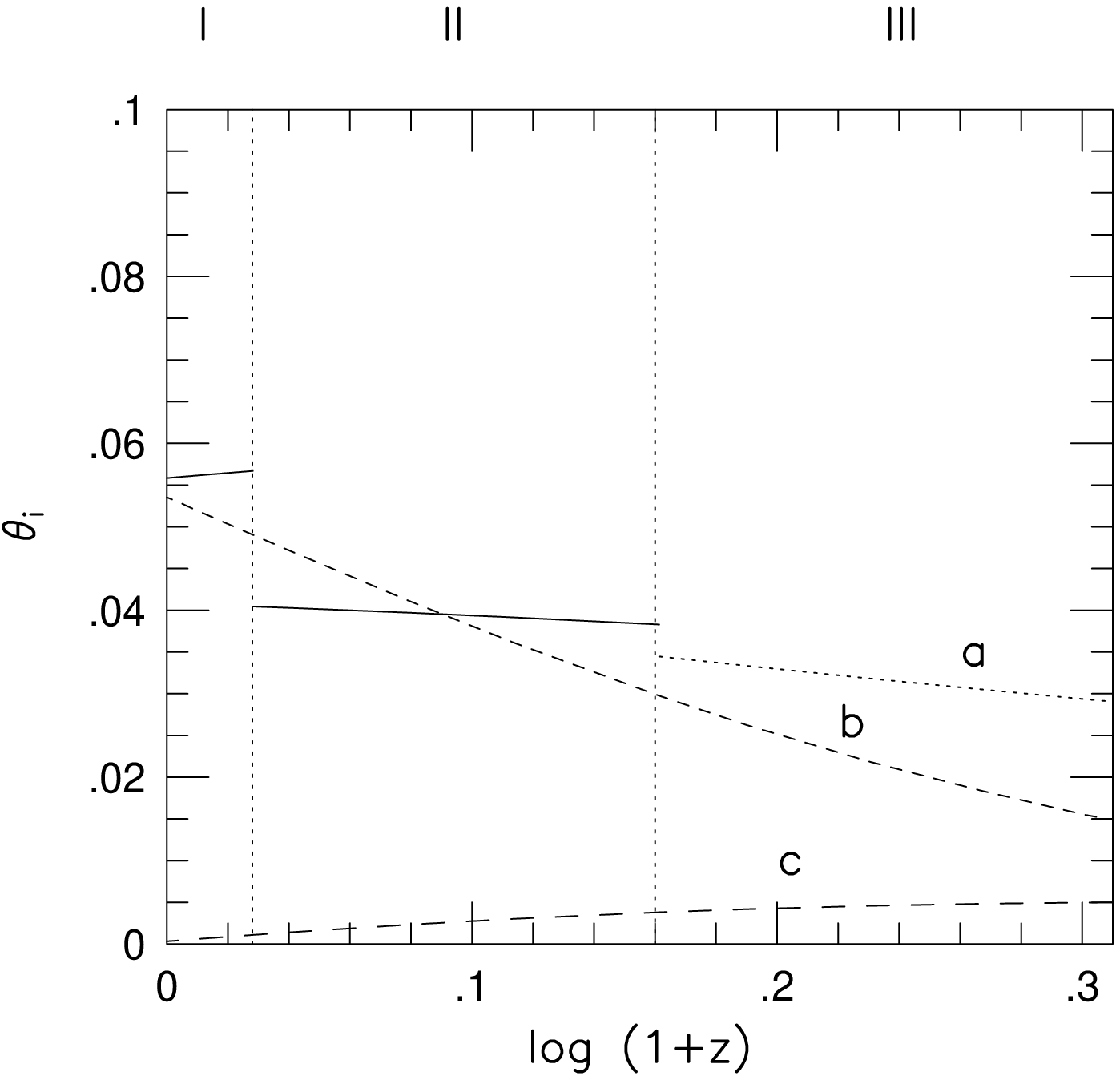}
    \end{center}
  \end{minipage}
 \end{tabular}
\vspace{-4cm}
 \caption{\label{fig:void0} The $z$-dependence of first and second
order temperature fluctuations in case 1(left) and in case 2 (right) 
for photons passing 
through the center of a compensated spherical void at $z < 1$. 
Solid curves denote $\theta_{I}$ and $\theta_{II}$ and the curve $a$
denotes $\theta_{III}$. The curves $b$ and $c$ denote
$\theta^{(1)}_\Lambda$ and $\theta^{(2)}_\Lambda$, respectively.  The
dotted vertical lines denote the boundaries at $z = 0.067$ and $z =
0.45$. We adopted $(\epsilon_{m0})_c = (\epsilon_{Im0})_c = -0.37$, for 
which $-\theta_{III}$ is comparable with $\theta_I$ and $\theta_{II}$.}  
\end{figure}

\begin{figure}[htbp]
 \begin{tabular}{cc}
  \begin{minipage}{0.5\hsize}
   \begin{center}
    \includegraphics[width=11cm,clip]{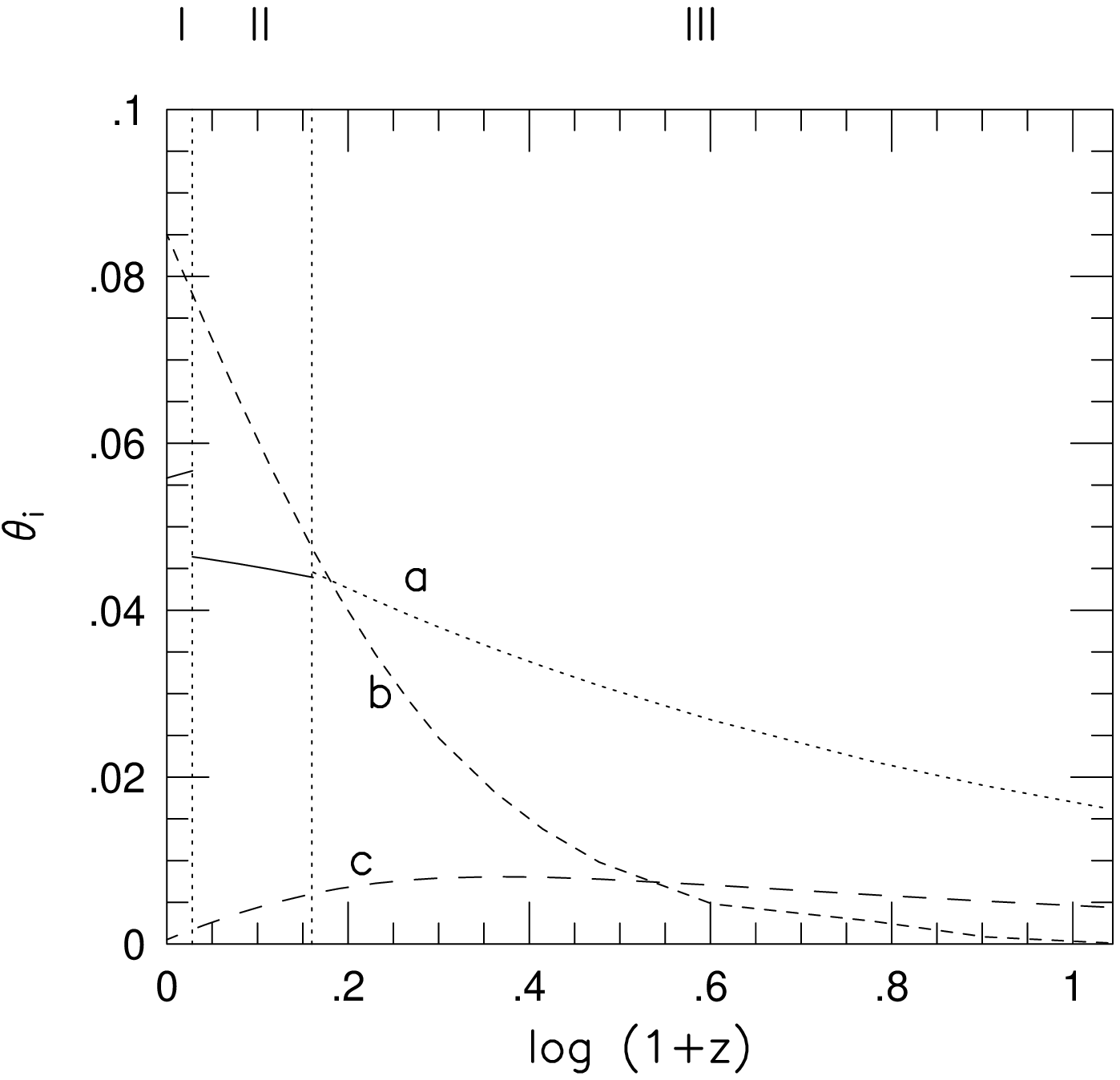}
   \end{center}
  \end{minipage}
  \begin{minipage}{0.5\hsize}
   \begin{center}
    \includegraphics[width=11cm,clip]{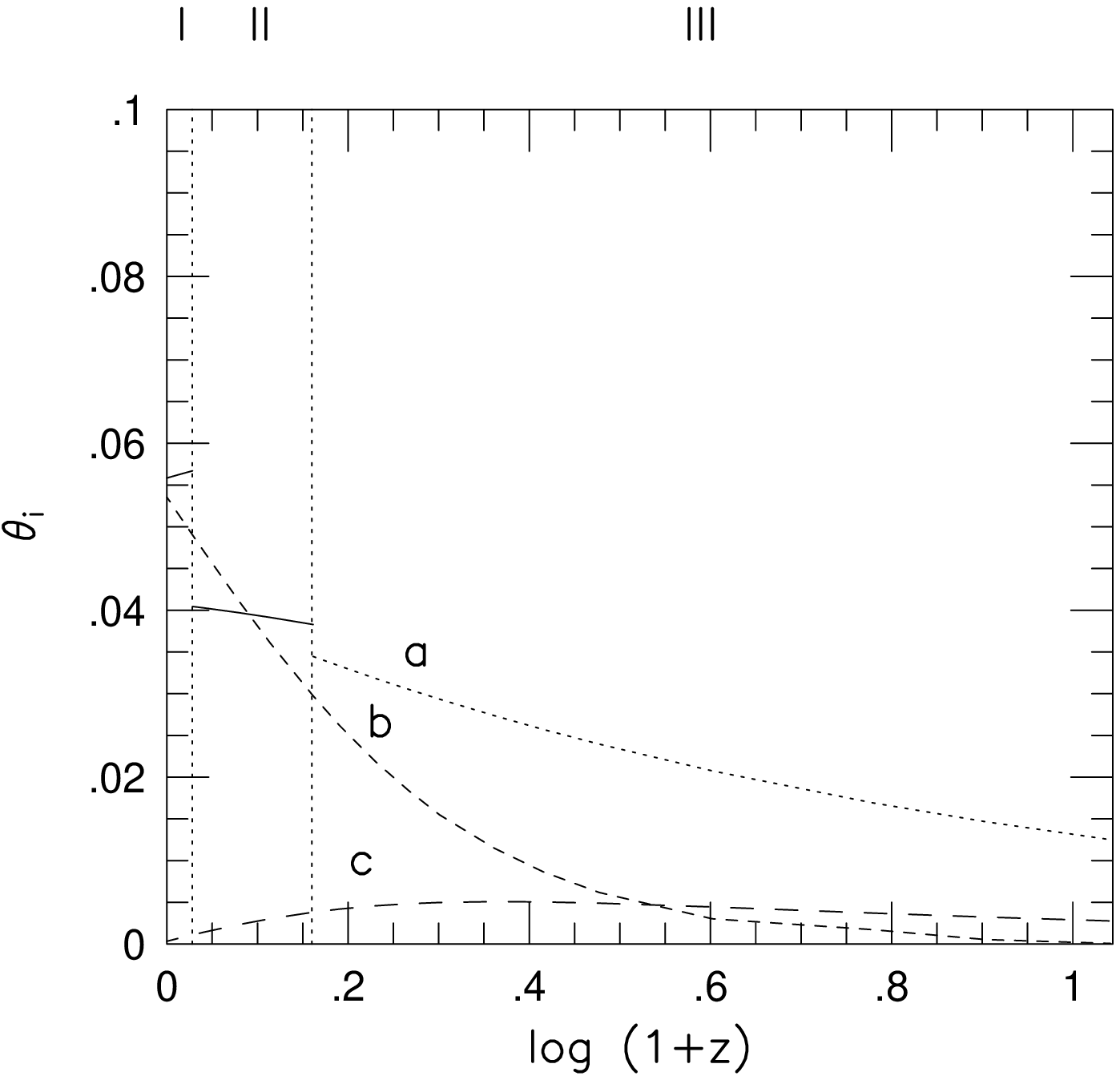}
    \end{center}
  \end{minipage}
 \end{tabular}
\vspace{-4cm}
 \caption{\label{fig:void2} The $z$-dependence of first and second
order temperature fluctuations in case 1(left) and in case 2 (right) 
for photons passing 
through the center of a compensated spherical void at $z < 10$. 
Solid curves denote $\theta_{I}$ and $\theta_{II}$ and the curve $a$
denotes $\theta_{III}$. The curves $b$ and $c$ denote
$\theta^{(1)}_\Lambda$ and $\theta^{(2)}_\Lambda$, respectively.  The
dotted vertical lines denote the boundaries at $z = 0.067$ and $z =
0.45$. We adopted $(\epsilon_{m0})_c = (\epsilon_{Im0})_c = -0.37$, for 
which $-\theta_{III}$ is comparable with $\theta_I$ and $\theta_{II}$.}  
\end{figure}

In Fig. \ref{fig:void0}, we show the behaviors of
$\theta_I, \theta_{II}$ and $-\theta_{III}$ in the interval $0 < z <
1$, in cases 1 and 2, respectively. In Fig. \ref{fig:void2}, in a similar manner, we show the behaviors of
$\theta_I, \theta_{II}$ and $-\theta_{III}$ in the interval $0 < z <
10$, in cases 1 and 2, respectively. In these figures we adopted 
$(\epsilon_{m0})_c = (\epsilon_{Im0})_c = -0.37$, for 
which $-\theta_{III}$ is comparable with $\theta_I$ and $\theta_{II}$,
and $\theta^{(1)}_\Lambda$ is smaller than $-\theta^{(2)}_\Lambda$ for
$z > 2.5$.   For comparison,
$\theta^{(1)}_\Lambda$ and $-\theta^{(2)}_\Lambda$ are also 
shown in them. Here the second-order
quantities are multiplied by $-1$, 
because they are negative definite and we use only positive
quantities in the figures. From these figures, we can see that $\theta_I,
\theta_{II}$ and $\theta^{(1)}_\Lambda$ are 
comparable in the regions I and II, though their behaviors are
different. 

 It is found that $|\theta^{(2)}_\Lambda|$ is smaller than
$|\theta_{III}|$, though both quantities are of second-order. 
This reflects the strong dependence on the Hubble constants
(cf. Eq.(\ref{eq:r10}) and Eq.(\ref{eq:r12})) 
and the $\Lambda$-dependence of the second-order ISW
effect which was studied in the previous paper\cite{ti}. 
  
As a result we find the following common features in cases 1 and 2
from these figures. 

\noindent (1) In the inner regions I and II, $\theta_I, \theta_{II}$
and $\theta^{(1)}_\Lambda$ are comparable, 
irrespective of $r_1, (\epsilon_{Im0})_c$ and $(\epsilon_{m0})_c$,
though $\theta_I$ and $\theta_{II}$ in case 1 seem to be larger by a
factor $\sim 1.5$ than $\theta_I$ and $\theta_{II}$. 

\noindent (2) In the outer region near the wall of $z = 0.45$,
$|\theta_{III}|$ is roughly comparable with $\theta^{(1)}_\Lambda$ for
$|(\epsilon_{Im0})_c| = 0.37$ and it is smaller or larger 
than $\theta^{(1)}_\Lambda$ for $|(\epsilon_{Im0})_c| <$ \ or \
$>0.37$, respectively. Far outside the wall, $|\theta_{III}|$ is
larger than $\theta^{(1)}_\Lambda$.

\noindent (3) Since $|\epsilon_{m0}|$ is $\approx 1$ for 
perturbations with a size $L \approx 10 h^{-1}$Mpc \ ($H_0 = 100h$
km/s/Mpc), $\theta_{III}$ is extremely large or negligible compared
with $\theta^{(1)}_\Lambda$ for perturbations with
$L \ll 10 h^{-1}$ or $\gg 10 h^{-1}$Mpc, respectively. 

\noindent (4) First-order quantities $\theta_I, \theta_{II}$ and
$\theta^{(1)}_\Lambda$ are 
positive/negative for a cluster/void with $(\epsilon_{Im0})_c$
and $(\epsilon_{m0})_c$, respectively, while second-order quantities
$|\theta_{III}|$ and $\theta^{(2)}_\Lambda$ are negative
definite. Therefore, in the concordant model, 
the expected amplitude is larger for voids than clusters. Such an asymmetry
is not expected in the outer region in our inhomogeneous model.

\noindent (5) At epochs of $z < z_c (= 2.5$ for $|(\epsilon_{m0})_c|
\approx 0.37)$, $|\theta^{(2)}_\Lambda| < 
|\theta^{(1)}_\Lambda|$, so
that the temperature fluctuations in the flat-$\Lambda$ models
have different signs for a cluster/void with a density contrast
 $\epsilon_{m0}$. For $z > z_c$, however, the second-order ISW 
effect is dominant also in the flat-$\Lambda$ model. 
Therefore, the temperature
fluctuations in the flat-$\Lambda$ models for $z>z_c$ is 
negative definite as in the outer region of our inhomogeneous models.


\section{Concluding remarks}
\label{sec:level4}

In this paper we studied the first-order and second-order ISW effect
in our anti-Copernican inhomogeneous model with underdense
regions in comparison with that in the concordant flat-$\Lambda$
model. We found that a distinct feature appears at the outer region at
moderate redshifts. In the concordant model,  
the expected amplitude of temperature anisotropy 
is larger for voids than clusters whereas such an asymmetry
is not expected in the outer region in our inhomogeneous model.
We showed, moreover,
that the first-order ISW effect in the inner regions of our models
is comparable with that in the flat-$\Lambda$ model, and that the
second-order ISW effect in the outer region depends on the amplitude
$\epsilon_{m0}$ of density perturbations.   
The ISW effect due to perturbations on scales larger than $100 h^{-1}$ Mpc 
with a density contrast $|\epsilon_{m0}|<0.37$ in the outer region is
negligible. On the other hand, in the inner region, 
no ISW effect appears due to 
perturbations on scales larger than the radius of the inner region. In our 
inhomogeneous model with underdense regions, 
the ISW effect does not contribute to the low-multipole 
components of CMB anisotropies\cite{olv,cont} in accord with 
the assertion proposed by Hunt and Sarker\cite{hunt}, while in the
flat-$\Lambda$ model, the contribution from the ISW effect due to
large-scale linear perturbations is significant. 

The observed correlation between the CMB sky with the 
large-scale structure is usually interpreted as the evidence
of the cosmological constant $\Lambda$, which causes the first-order
ISW effect\cite{bou,turok}. Recently, moreover,  
hot and cold spots on the CMB sky associated with super-structures
(with $z \sim 0.5$) in SDSS Luminous Red Galaxy catalog were measured
by Granett et al.\cite{granett} and the consistency with the ISW
effect in the flat-$\Lambda$ models was shown. It should be
noted, however, that they may be brought in principle by the
first and second-order ISW effect also in our models with 
underdense regions, as we showed in the present paper. Therefore,  the
observational  
evidence for the existence of small-scale ISW effect for light paths 
through clusters of galaxies, superclusters and supervoids may support
not only the flat-$\Lambda$ model, but also our models with 
underdense regions.
 
In order to make a clear distinction between the two models,
it is better to compare the overall amplitudes of temperature fluctuations 
associated with a void(negative density) with those associated with 
a cluster(positive density). As we have seen, for $0.45 < z < z_c$
(which depends on 
$\epsilon_{m0}$), the amplitudes for quasi-linear voids are larger
than those for 
quasi-linear clusters in the concordant model due to the second-order
effect, while such an asymmetry cannot be
expected in our inhomogeneous model since there is no 
first-order effect in the outer region\cite{ti,si}.

\appendix
\section{Definitions of $p(\eta), q(\eta), \zeta_1(\eta)$ and
$\zeta_2(\eta)$}
 
$P(\eta)$ satisfies
\begin{equation}
\label{eq:r13}
P'' + {2a' \over a} P' -1 = 0
\end{equation}
and its solution is expressed as
\begin{eqnarray}
  \label{eq:r14}
P(\eta) &=& -{2 \over 3\Omega_{m0}}{a}^{-3/2}
[\Omega_{m0}+\Omega_{\Lambda 0} {a}^3]^{1/2}
\int^{{a}}_0 d \tilde{a}
\tilde{a}^{3/2}[\Omega_{m0}+\Omega_{\Lambda 0} \tilde{a}^3]^{-1/2} +
{2 \over 3\Omega_{m0}}{a}, \cr
\eta &=& \int^{{a}}_0 
d\tilde{a} \tilde{a}^{-1/2}[\Omega_{m0}+\Omega_{\Lambda 0}
\tilde{a}^3]^{-1/2}.
\end{eqnarray}
The functions $\zeta_1$ and $\zeta_2$ are defined as
\begin{eqnarray}
  \label{eq:r15}
\zeta_1  &=& {1 \over 4}P \Bigl(1 - {a' \over a}P'\Bigr), \cr
\zeta_2  &=& \Big\{{1 \over 21} {a' \over a}\Bigl(PP' - {1 \over
6}Q'\Bigr) - {1 \over 18}\Bigl[P + {1 \over 2}(P')^2\Bigr] \Big\},
\end{eqnarray}
where $Q(\eta)$ is satisfies
\begin{equation}
  \label{eq:r16}
Q'' + {2a'\over a} Q' = - \Bigl[P - {5\over 2}(P')^2 \Bigr].
\end{equation}
%



\end{document}